\documentclass[onecolumn,amsmath,showkeys,amssymb]{revtex4}

\usepackage{graphicx,color}

\usepackage{amsbsy}
\usepackage{amsthm,mathrsfs,amsopn}

\begin{document}

\title{Delay induced Turing-like waves for one species reaction--diffusion model on a network}
\author{Julien Petit$^1$, Timoteo Carletti$^1$, Mabor Asslani$^1$, Duccio Fanelli$^2$}

\affiliation{1. naXys, Namur Center for Complex Systems, University of Namur, rempart de la Vierge 8, B 5000 Namur, Belgium\\
2. Dipartimento di Fisica e Astronomia, University of Florence,INFN and CSDC, Via Sansone 1, 50019 Sesto Fiorentino, Florence, Italy}

\begin{abstract} 
A one species time--delay reaction-diffusion system defined on a complex networks is studied.  
Travelling waves are predicted to occur as  follows a symmetry breaking instability of an homogenous stationary stable solution, subject to an external non homogenous perturbation. These are generalized Turing-like waves that materialize in a single species populations dynamics model, as the unexpected byproduct of the imposed delay in the diffusion part.  Sufficient conditions for the onset of the instability are mathematically provided by performing a linear stability analysis adapted to time delayed differential equation. The method here developed exploits the properties of the Lambert W-function. The prediction of the theory are confirmed by direct numerical simulation carried {out} for a modified version of the classical Fisher model, defined on a Watts-Strogatz networks and with the inclusion of the delay.
\end{abstract}

\maketitle

\vspace{0.8cm}


Collective dynamics spontaneously emerge in a vast plethora of physical systems and often play a role of paramount importance for the efficient implementation of dedicated functions. Travelling waves are among the most studied phenomena for their ubiquitous and cross-disciplinary interest. Periodic traveling waves are for example encountered when describing self-oscillatory and  excitable systems in different realms, from chemistry to biology, passing through physics.  The Fisher \cite{Fisher1930,Fisher1937} equation, introduced to characterize the spatial spread of an advantageous allele, defines the paradigmatic arena for addressing the peculiarities of travelling wave solutions in a reaction diffusion system {in a spatial continuous domain}. This is a one species population dynamic model, which assumes a logistic rule of replication and growth. The microscopic entities belonging to the scrutinized population can also delocalized in space, as follows a standard diffusion mechanism. The interplay between the aforementioned processes yields stable travelling wave solutions, which are selectively generated starting from a special class of initial conditions \cite{kolmogorov}. More generally, it is however interesting to speculate on the possibility for a system to yield self-organized collective patterns of the travelling wave type, as follows a symmetry breaking instability, seeded by diffusion. Starting from a homogeneous solution subject to a tiny, non homogeneous, initial perturbation, a reaction diffusion system can destabilize via a dynamical instability, identified by Alan Turing in a seminal work \cite{turing1952}. The Turing instability, as the process is nowadays called, can drive the emergence of non linear stationary stable patterns, if at least two species, the activator and inhibitors, are diffusiong and mutually interacting in the embedding environment. Alternatively, the Turing mechanims can instigate travelling wave solutions, provided at least three species, one of which mobile, are assumed to interact via apt non linear couplings. For a reaction diffusion system hosted on a discrete heterogeneous spatial support, namely a network, the instability can eventually set in for a two species model, if just the inhibitor is allowed to crawl from one node to its adjacent neighbors \cite{Cantini2013}. Also in this case, three coupled species are the minimal request for a travelling wave to rise from a stochastic perturbation of an initial homogeneous stationary stable state. 

Starting from these premises, the aim of this Letter is to tackle the above problem under a radically different angle 
and thus introduce the simplest mathematical setting for which travelling wave solutions are generated, on a network, as a symmetry breaking instability of the Turing type. To anticipate our finding, we will prove that a one species model endowed with a delay in the diffusion can produce the sought instability. The analysis holds in general, but to demonstrate our conclusion we shall refer to a Fisher equation defined on a complex networks and modified with the inclusion of the delay term. 

The usage of time--delay differential equations (DDEs) defined on complex networks ~\cite{AtayJost2004,PonceMM2009,HofenerSG2011} is nowadays very popular, from pure to applied sciences~\cite{Niculescu2001,Richard2003}. Delays are for instance introduced to model finite communication or displacement time of quantities across network links. The imposed time-delay can non trivially  interfere with the reactive dynamics, taking place on each node of the graph, thus resulting in unexpected emergent properties.
For example, the classical paper~\cite{SchusterWagner} deals with time-delayed system made of two coupled phase oscillators and demonstrates the existence of multistability of synchronized solutions: an invariant manifold exists which attracts all the solutions of the system, yielding global oscillatory phenomena. Since this pioneering contribution, the subject has gained lot of attention and several contribution have been reported, assuming linear systems~\cite{JirsaDing2004}, coupled oscillators~\cite{Kim1997,YeungStrogatz1999,EarlStrogatz2003}, oscillations death~\cite{Reddy1998,Atay2003,Dodla2004} and oscillations control~\cite{RosenblumPikovsky2004}. Theory has been fruitfully applied to tackle real problems, see for instance~\cite{Herrero2000,Kozyreff2000,Reddy2000,Takamatsu2000,Vladimirov2003,Takamatsu2004,RosenblumPikovsky2004b}.

The work of this Letter moves from this reference context to build an ideal bridge with the Turing-like theory of pattern formation on complex networks. The discrete Laplacian operator, that encodes for the diffusion of the mobile population, incorporates a delay term. Our analysis of the DDEs exploits the properties of the Lambert W-function~\cite{CorlessEtAl1996,AslUlsoy2003} {thus} the solution of the linearised equation can be given in closed analytical form. Furthermore, we have full access to the associated eigenvalues and their dependence on the involved parameters is made explicit. This allows us to go beyond the technique based on the computation of the Hopf bifurcation, namely to determine the parameters for which the eigenvalue with the largest real part passes through a pure imaginary value.

We consider an undirected connected network composed by $n$ nodes and assume one species to diffuse, from node to node, via the available links.  Reactions 
also take place on each node, as dictated by a specific non-linear function $f$ of the local species concentration. Let us denote by $x_i(t)$ the species concentration on node $i$, at time $t$. Then its time evolution is governed by
\begin{equation}
\label{eq:main0}
\dot{x}_i(t)=f(x_i(t-\tau_r))+D \sum_j L_{ij} x_j(t-\tau_d)\, ,
\end{equation}
where $D$ stands for the diffusion coefficient, $L_{ij}=G_{ij}-k_i\delta_{ij}$ is the Laplacian matrix of the network whose adjacency matrix is given by $G$ and $k_i=\sum_j G_{ij}$ identifies the degree of the $i$-th node. $\tau_r$ is the delay involved in the reaction occurring at each node, while $\tau_d$ is the delay due to the displacement across nodes. For the sake of simplicity we hereby assume the delay to be independent from the link indexes~\footnote{A more general description can be provided by taking into account the possibility that each link introduces a different delay~\cite{PonceMM2009}}. As mentioned earlier, it is well known that the above one species reaction-diffusion system cannot exhibit Turing-like instability, in the limiting case for $\tau_r=\tau_d=0$. As we shall argue, the introduction of a finite delay
$\tau_r=\tau_d=\tau>0$, will significantly alter this conclusion. 

To proceed in the analysis we assume a stable homogeneous equilibrium, $x_i(t)=\hat{x}$ for all $i=1,\dots,n$ and $t\geq 0$ and look for sufficient conditions to destabilize such equilibrium, as follows the introduction of a non homogeneous perturbation, which in turn activates the diffusion part. To determine the preliminary conditions that have to be met for the homogeneous equilibrium to be stable, we linearize system~\eqref{eq:main0} with $D=0$, around $\hat{x}$, and recall that $\tau_r=\tau>0$. Let $A = f^{\prime}(\hat{x})$, then the characteristic equation reads $\lambda  = A e^{-\lambda \tau}$ whose solutions are:
\begin{equation}
\lambda_k = \frac{1}{\tau} W_k(\tau A),\quad k\in \mathbb{Z}\, ,
\end{equation}
$W_k$ being the $k$--th branch of the Lambert W--function~~\cite{CorlessEtAl1996}. To guarantee the needed stability of the homogeneous equilibrium $\hat{x}$, one has to require that $(A,\tau)$ and an integer $k$ exist for which $\Re\lambda_k(A,\tau)<0$.

\begin{figure*}[ht!]
\begin{center}
\includegraphics[width=7cm]{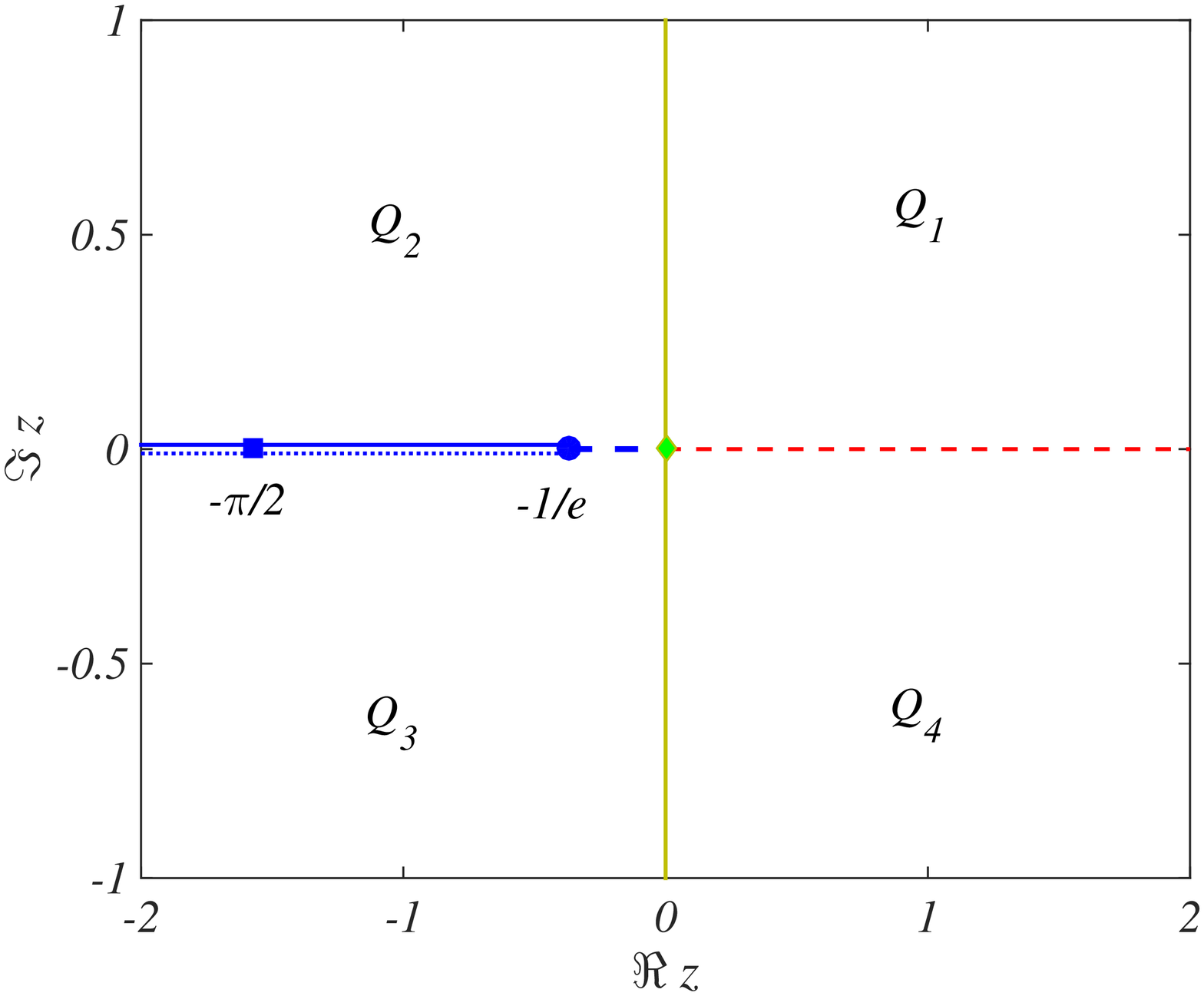}\quad\includegraphics[width=7cm]{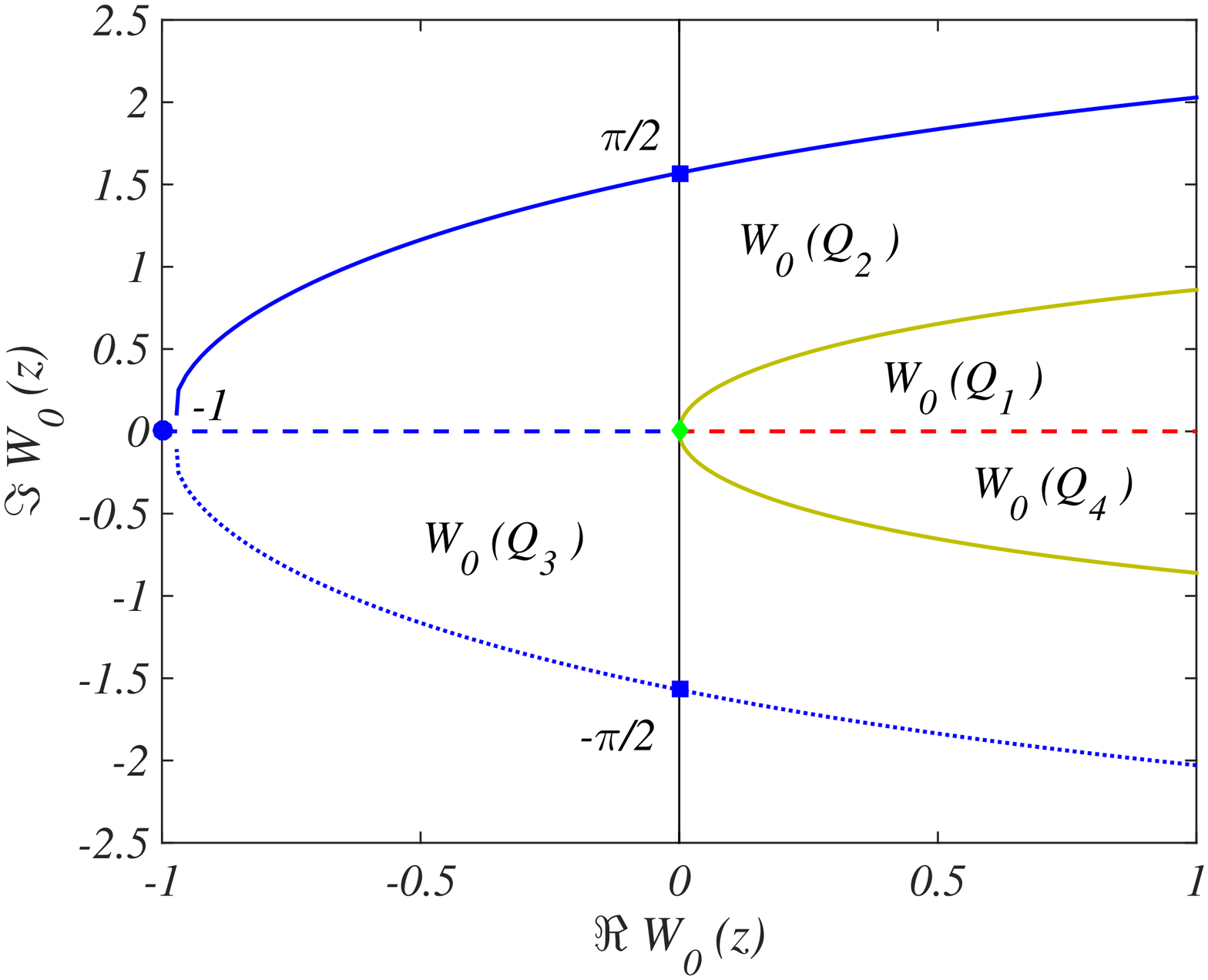}
\end{center}
\caption{The Lambert $W$ function: principal branch $W_0(z)$. Left panel: in the complex plane $z\in\mathbb{C}$, we represent the upper part of the branch cut $\{z : - \infty < \Re z \leq -e^{-1}\, , \, \Im z=0^+\}$ by a solid line and the lower part of the branch cut $\{z : - \infty < \Re z \leq -e^{-1}\, , \, \Im z=0^-\}$ by a dashed line; the circle denotes the point $(-1/e,0)$ while the square refers to $(-\pi/2,0)$. Right panel: the complex plane $w\in\mathbb{C}$, where $w=W_0(z)$. The solid blue line is the image of the upper part of the branch cut $\Im w \mapsto \Re w=-\Im w\, \mathrm{co}\!\tan \Im w$ for $0<\Im w <\pi$, while the dashed blue line is the image of the lower branch cut via $W_0$, $\Im w \mapsto \Re w=-\Im w \,\mathrm{co}\!\tan \Im w$ for $-\pi<\Im w <0$. The circle at coordinates $(-1,0)$ is the image of the point $(-1/e,0)$ and the squares positioned at $(0,\pi/2)$, respectively at $(0,-\pi/2)$, are the image of the point of coordinates $(-\pi/2,0^+)$, respectively $(-\pi/2,0^-)$. The red dashed line is the image of the positive real axis while the green curved line is the image of the imaginary axis $\Re z=0$. Observe that if $\Im z=0$ and $-\pi/2<\Re z < -1/e$, then $-1\leq\Re W_0(z)<0$, $\lvert\Im W_0(z)\rvert<\pi/2$ and $\Im W_{0}(z)\neq 0$ (blue solid and dotted lines), if $\Im z=0$ and $-1/e\leq \Re z\leq 0$, then $-1\leq\Re W_0(z)\leq 0$ and $\Im W_0(z)=0$ (blue dashed line), if $\Im z=0$ and $\Re z < -\pi/2$, then $\Re W_0(z)>0$ and $\lvert\Im W_0(z)\rvert>\pi/2$ (blue solid and dotted lines) and if $\Im z=0$ and $\Re z> 0$, then $\Re W_0(z)>0$ and $\Im W_0(z)=0$ (red dashed line).}
\label{fig:lambertW}
\end{figure*}

The Lambert W--function is the complex multivalued function of the complex variable $z\in\mathbb{C}$ defined to be solution of the equation $z=W(z)e^{W(z)}$. It has infinitely many branches~\cite{CorlessEtAl1996} denoted by $W_k(z)$, for $k\in\mathbb{Z}$; among them $W_0(x)$ - the principal branch - is obtained by restricting $z$ to lie on the real axis, more precisely on $\Re z\in(-e^{-1},+\infty)$, and with the constraint $W_0(z)\geq -1$. Hence the branch cut of $W_0$ is defined by $\{z : - \infty < \Re z \leq -e^{-1}\, , \, \Im z=0\}$. Let us observe that $k=0$ and $k=-1$ are the only branches for which the Lambert W--function can assume real values. For all remaining $k$, $\Im W_k(z)\neq 0$ for all $z$ (see Fig.~\ref{fig:lambertW}).

Roughly speaking, $W_0$ bends the $z$ plane (cut along $\Re z<-e^{-1}$) into a parabolic like domain in the $w$ plane, whose boundary curves $\Im w \mapsto \Re w=-\Im w \,\mathrm{co}\!\tan \Im w$ (blue solid and dotted curves in Fig.~\ref{fig:lambertW}) are bounded by $\pi$ and $-\pi$. Moreover $W_0$ satisfies the following relevant condition (Lemma 3 of~\cite{ShinozakiMori2006})
\begin{equation}
\label{eq:wkw0}
\forall z\in \mathbb{C}\quad : \quad \max_{k\in \mathbb{Z}}\Re W_k(z)=\Re W_0(z)\, .
\end{equation}
  
The previous Eq.~\eqref{eq:wkw0} allows to restate the stability condition of the homogeneous equilibrium as follows:
\begin{equation}
\exists (A,\tau) \text{ such that } \Re W_0(\tau A)<0\, .
\end{equation}
Hence, from the properties of the Lambert W--function represented in Fig.~\ref{fig:lambertW}, this amounts to require
\begin{equation}
\label{eq:main1stabCond}
-\frac{\pi}{2}<\tau A<0\, .
\end{equation}

We now turn to considering the effect of a non homogeneous perturbation, superposed to the postulated homogenous equilibrium.
This implies studying the full system~\eqref{eq:main0}, with $\tau_r=\tau_d=\tau>0$ and $D>0$. Linearising as before, we get
\begin{equation}
\label{eq:main1perturbation}
\dot{\delta x}_i(t) = A \delta x_i(t-\tau)+D\sum_j L_{ij}\delta x_j(t-\tau)\, .
\end{equation}
The Laplacian matrix has a complete set of orthonormal eigenvectors $\phi^\alpha$, associated to the topological eigenvalues $0=\Lambda^1>\Lambda^2\geq\ldots\geq\Lambda ^n$, $\sum_j L_{ij}\phi^\alpha_j=\Lambda^\alpha\phi^\alpha_i$, for $\alpha=1,\ldots,n$.  Using this basis to decompose $\delta x_i(t)$ and employing once again the ansatz of exponential growth
\begin{equation}
\delta x_i(t) = \sum_\alpha c_\alpha e^{\lambda_\alpha t} \phi^\alpha_i,\quad \mbox{for all}\quad i=1,\ldots,n\, ,
\end{equation}
we eventually get from~\eqref{eq:main1perturbation} the characteristic equation 
\begin{equation}
\lambda_\alpha -(A+D\Lambda^\alpha )e^{-\lambda_\alpha \tau}=0\, .
\end{equation}
That is $\tau \lambda_\alpha e^{\tau \lambda_\alpha}=\tau (A+D\Lambda^\alpha)$, whose solutions are:
\begin{equation}
\lambda_{\alpha,k} = \frac{1}{\tau}W_k\left(\tau(A+D\Lambda^\alpha)\right),\quad k\in \mathbb{Z}\text{ and $\alpha=1,\dots, n$.}
\end{equation}

Symmetry breaking instabilities seeded by diffusion can set in if at least a pair $k$ and  $\alpha$ exists such that $\Re \lambda_{\alpha,k}>0$, provided $-\frac{\pi}{2}<\tau A<0$ (the homogeneous equilibrium must be stable as a prerequisite of the analysis)~\footnote{Remark that $\alpha >2$: in fact, for $\alpha=1$ we have $\Lambda^1=0$ and thus, by assumption, $\Re \lambda_{1,k}=\Re W_k(\tau A)<0$.}. When a bounded family of $k$ exist for which $\Re \lambda_{\alpha,k}>0$, the one with the largest real part dominates the instability and shapes the emerging pattern. Recalling again Eq.~\eqref{eq:wkw0}, we finally get the following sufficient condition for the onset of Turing-like instability in a one species DDE of the general type \eqref{eq:main0}:

\begin{equation}
\label{eq:TPcond}
\exists \alpha \in [2,\dots , n]\quad \text{such that }\frac{1}{\tau} \Re W_0\left(\tau (A+D\Lambda^\alpha)\right)>0\, . 
\end{equation}

Exploiting the properties of the Lambert W--function represented in Fig.~\ref{fig:lambertW}, Eq.~\eqref{eq:TPcond} is satisfied whenever~\footnote{It is worth emphasizing that, in principle, one could also satisfy Eq.~\eqref{eq:TPcond} with $\tau (A+D\Lambda^\alpha)>0$. This alternative condition is however not compatible with the stability request~\eqref{eq:main1stabCond} and the negativeness of $\Lambda^\alpha$ for all $\alpha$.}  there exists $\bar{\alpha}>1$  such that
\begin{equation}
\label{eq:turing1}
\tau D\Lambda^{\bar{\alpha}} < -\frac{\pi}{2} -\tau A\, .
\end{equation}
Observe also that for such $\bar{\alpha}$, $\Im \lambda_{0,\bar{\alpha}}\neq 0$. Hence, the instability materialize in the appearance of travelling waves. 
Since $\Re W_0(x)$ is increasing, for decreasing real $x<-\pi/2$, the dispersion relation attains its maximum at $\alpha=n$, the Laplacian eigenvalues being ordered for decreasing real parts. Hence, the Turing-like instabilities cannot develop if $\tau D\Lambda^{n} > -\frac{\pi}{2} -\tau A$. Summing up, travelling waves are expected to develop, for a fixed network topology, and sufficently large values of $\tau$ and $D$. Conversely, for a fixed choice of the parameters ($\tau$, $D$ and $A$), one should make the network big and so force $\Lambda^n$ to be large enough, in absolute value. 

As our goal is to determine the minimal model for which the aforementioned instability sets in, we now consider the reduced case for 
$\tau_r=0$ and $\tau_d=\tau>0$, that is the delay is only associated to the diffusion part. For $D=0$ the homogeneous equilibrium $\hat{x}$, is stable if and only if $A<0$. 
By repeating the procedure highlithed above we end up with the following characteristic equation:
\begin{equation}
\lambda_\alpha - A - D \Lambda^\alpha e^{-\lambda_\alpha \tau }=0\, , 
\end{equation}
whose solution reads
\begin{equation}
\lambda_{\alpha,k} = \frac{1}{\tau}W_k(\tau D \Lambda^\alpha e^{-\tau A})+A, \quad k \in \mathbb{Z}\, . 
\end{equation}
The equilibrium $\hat{x}$ is hence destabilised by diffusion if there exist $\alpha>1$ for which
\begin{equation}
\label{eq:main3instabCond}
\Re W_0(\tau D \Lambda^\alpha e^{-\tau A})>-A\tau\, . 
\end{equation}
Let us observe that $-A\tau$ is positive and $\tau D \Lambda^\alpha e^{-\tau A}$ is negative. Moreover, we already remarked that $\Re W_0(x)$ is positive and increasing for decreasing negative $x$. Hence, a critical $x_c<0$ exists~\footnote{See section A of the Appendix the explicit computation of $x_c$.} for which $\Re W_0(x_c)=-A\tau$ and Eq.~\eqref{eq:main3instabCond} is satisfied for all $\tau D \Lambda^\alpha e^{-\tau A}<x_c$.  Travelling waves are hence predicted to manifest,  for sufficiently large  $\Lambda^{\bar{\alpha}}$, in absolute value. We again stress that $\Im \lambda_{0,\bar{\alpha}}\neq 0$. 

To complete the general discussion we consider the dual problem, where $\tau_r=\tau>0$ and $\tau_d=0$. As already observed, the homogenous equilibrium $\hat{x}$ is stable if $-\pi/2<\tau A<0$. The characteristic equation associated to this problem can be cast in the form:
\begin{equation}
\lambda_\alpha - A e^{-\lambda_\alpha\tau}-D\Lambda^\alpha =0\, ,
\end{equation}
whose solution is
\begin{equation}
\lambda_{\alpha,k} = D\Lambda^\alpha + \frac{1}{\tau}W_k(\tau A e^{-\tau D \Lambda^\alpha}),\quad k\in\mathbb{Z}\, .
\end{equation}
One can prove~\footnote{See section B of the appendix for a rigorous proof of the claim.}  that $D\Lambda^\alpha + \frac{1}{\tau}W_0(\tau A e^{-\tau D \Lambda^\alpha})<0$ for all $\Lambda^\alpha$, $\alpha>1$. We are consequently led to conclude that the stable homogeneous equilibrium cannot undergo a diffusion driven Turing-like instability, if the delay term is solely confined in the reaction part. 

\begin{figure*}[ht!]
\begin{center}
\includegraphics[width=7cm]{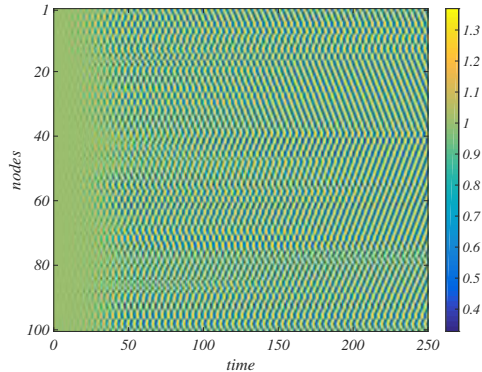}
\end{center}
\caption{Numerical solution of {the system under scrutiny}  with parameters $a=1.4$, $D=0.05$ and $\tau=1$. Initial conditions are set equal to $x_i(t)=1+\delta_i$ for all $t\in[-\tau,0)$ where $\delta_i$ are random  gaussian numbers drawn from $N(0,0.01)$. The underlying network is a Watts-Strogatz network~\cite{WattsStrogatz} made of $100$ nodes, with average degree $\langle k\rangle=6$ and probability to rewire a link $p=0.03$.}
\label{fig:patterns20042015_001}
\end{figure*}

\begin{figure*}[ht!]
\begin{center}
\includegraphics[width=7cm]{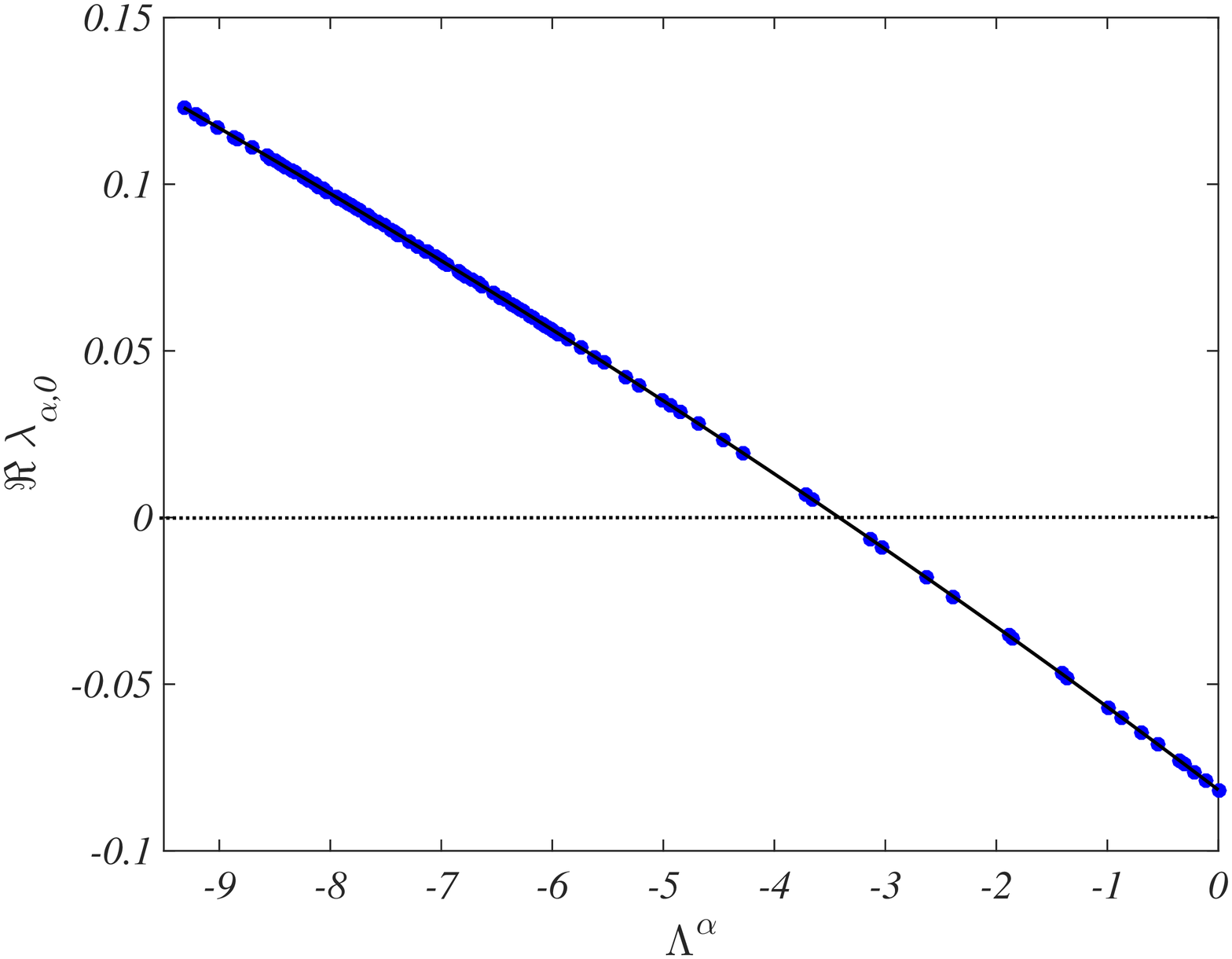}\quad\includegraphics[width=7cm]{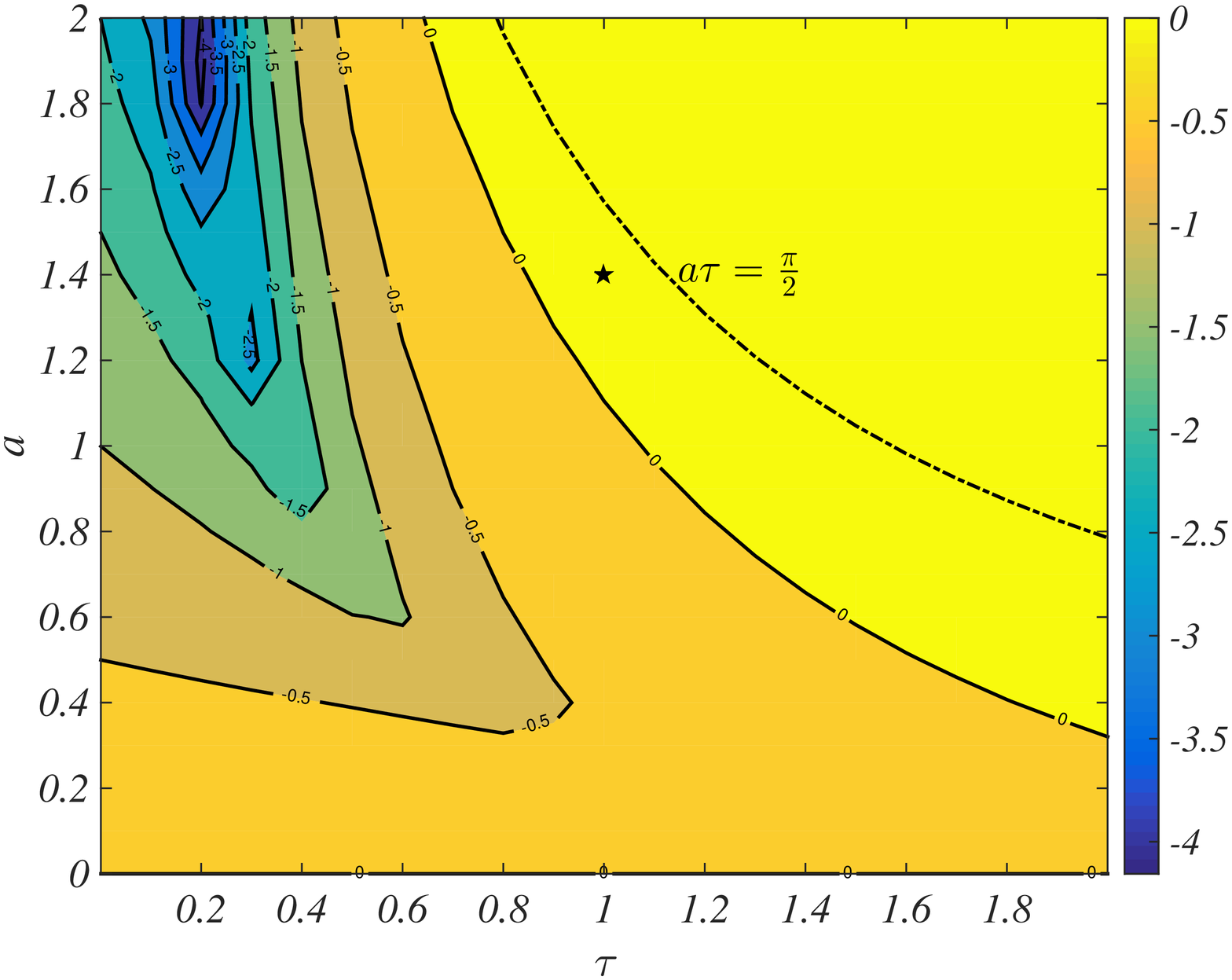}
\end{center}
\caption{Dispersion relation. Left panel $\Re\lambda_{\alpha,0}$ as a function of $\Lambda^{\alpha}$, for fixed $a$, $D$ and $\tau$, as specified in Fig.~\ref{fig:patterns20042015_001} and for the same Wattz-Strogatz network. Blue dots represent the values of $\lambda_{\alpha,0}$ computed for a given $\Lambda^{\alpha}$, while the solid black line is the continuous approximation. The horizontal dotted black line stands for the  $0$--th level. Right panel: $\max_{\alpha}\Re\lambda_{\alpha,0}$, as a function of $(a,\tau)$, for fixed $D=0.05$ using the same Wattz-Strogatz network as employed in Fig.~\ref{fig:patterns20042015_001}. The dash dotted curve $a\tau=\pi/2$ delineates the boundary (together with $a>0$ and $\tau>0$) of the stability region of the homogenous solution $\hat{x}=1$. Hence, all pairs $(a,\tau)$ for which $0<a\tau<\pi/2$ correspond to stable solution of the homogeneous equilibrium.  Turing-like instabilities are thus allowed to develop for all pairs $(a,\tau)$ inside such domain, for which $\max_{\alpha}\Re\lambda_{\alpha,0}>0$. The star refers to the setting of the simulation reported in Fig.~\ref{fig:patterns20042015_001}.}
\label{fig:RelDispAvsTauRDDD}
\end{figure*}

As an application of the previous theory, we take $f$ in Eq.  \eqref{eq:main0} to be logistic function $f(x)=ax(1-x)$, as in the spirit of the Fisher model. 
At variance with the Fisher equation~\cite{Fisher1930,Fisher1937}, we now imagine the species to be hosted on a discrete support, rather than a continuum segment. In the original Fisher scheme the emerging wave relates to the heteroclinic orbit of the system and requires a specific, step-like, initial profile. In our case the travelling wave will originate as follows a symmetry breaking instability of an initial randomic perturbation. To carry out the analysis,
we select the homogeneous equilibrium solution $\hat{x}=1$ and look after to its associated stability properties, as a function of $a$, $\tau$, $D$ and the network topology. 
Silencing the diffusion, $D=0$, the stability condition Eq.~\eqref{eq:main1stabCond} rewrites $0<a\tau <\frac{\pi}{2}$. In demonstrating our findings, we consider
a Watts-Strogatz network~\cite{WattsStrogatz} made of $100$ nodes, with average degree $\langle k\rangle=6$ and probability to rewire a link $p=0.03$. Other network topologies can be in principle assumed, returning similar qualitative conclusions. The chosen network is large enough so to have one eigenvalue for which the dispersion relation has positive real part (see left panel of Fig.~\ref{fig:RelDispAvsTauRDDD}): thus Turing-like waves do exist. In Fig.~\ref{fig:patterns20042015_001} we report the result of a numerical solution of {the system under scrutiny} obtained with a RK4 method adapted to deal with delay differential equation. The parameters are set to the values $a=1.4$, $D=0.05$ and $\tau=1$. The DDE should be complemented with the value of the function on the delay interval $[-\tau,0)$. In the spirit of a perturbation of the stable equilibrium, we decided to set $x_i(t)=1+\delta_i$ for all $t\in[-\tau,0)$ where $\delta_i$ are random Gaussian numbers drawn from $N(0,0.01)$. Observe that $a\tau=1.4<\pi/2$ and thus the equilibrium $\hat{x}=1$ is stable in absence of diffusion. On the other hand, one can clearly appreciate that after a transient period, patterns do manifest as stable oscillations around the solution $\hat{x}=1$.

In Fig.~\ref{fig:RelDispAvsTauRDDD} the dispersion relation are displayed. In the left panel the quantity
\begin{equation*}
\max_k\Re \lambda_{\alpha,k}=\Re \lambda_{\alpha,0}=\frac{1}{\tau}\Re W_0\left(-\tau a+\tau D\Lambda^{\alpha}\right)\, ,
\end{equation*}
is plotted as a function of $\Lambda^{\alpha}$, for fixed $a$, $D$ and $\tau$ (as specified in Fig.~\ref{fig:patterns20042015_001}), and for the same Watts-Strogatz network. On can clearly identify  several eigenvalues, for which $\Re \lambda_{\alpha,0}>0$. The right panel reports the maximum of the dispersion relation  as a function of the eigenvalues, i.e. $\max_{\alpha}\Re \lambda_{\alpha,k}$, versus $(a,\tau)$ for fixed $D=0.05$ and for the same Wattz-Strogatz network. The stability domain of the equilibrium without diffusion is bounded by $a>0$ ($\tau>0$, for physical reasons) and $a\tau<\pi/2$ (dash-dotted black curve). The Turing-like waves can emerge for all pairs $(a,\tau)$, inside such limited domain, for which $\max_{\alpha}\Re\lambda_{\alpha,0}>0$.

In conclusion, we have here shown that a one species time--delay reaction-diffusion system defined on a complex networks can exhibit 
travelling waves, as follows a symmetry breaking instability of an homogenous stationary stable solution, subject to an external non homogenous perturbation. 
These are Turing-like waves which emerge in a minimal model of single species population dynamics, as the unintuitive byproduct of the imposed delay.  Based on a linear stability analysis adapted to time delayed differential equation, we provided sufficient conditions for the onset of the instability, as a function key quantities, as the reaction parameters, the delay, the diffusion coefficient and the network topology. The wave possesses multiple fronts and persists in time, without fading away as it happens for the customary Fisher equation. The observation that Turing--like instability can originate for a one species model evolving on a heterogeneous graph, provided a delay is included in the transport term, enables us to significantly relax the classical constraints for the patterns to emerge and opens up the perspective for intriguing developments in a direction so far unexplored.

The work of J.P, T.C. and M.A. presents research results of the Belgian Network DYSCO (Dynamical Systems, Control, and Optimization), funded by the Interuniversity Attraction Poles Programme, initiated by the Belgian State, Science
Policy Office.

\appendix

\section{Turing-like instability when the delay is confined in the diffusion term.}
\label{sec:xc}

Let us consider the case where the delay is present only in the diffusion part ($\tau_d=\tau>0$ and $\tau_r=0$)
\begin{equation}
\label{eq:main0b}
\dot{x}_i(t)=f(x_i(t))+D \sum_j L_{ij} x_j(t-\tau)\, .
\end{equation}
To study the stability of the equilibrium $\hat{x}$ in this case we have to look for $\alpha>1$ for which 
\begin{equation}
\label{eq:tosolve}
\Re W_0(\tau D \Lambda^\alpha e^{-\tau A})>-A\tau\, . 
\end{equation}

Let $z=x+iy$ and $w=\xi+i\eta$, then the following relation is recovered once we impose $z=we^w$, namely $w=W_0(z)$:
\begin{equation}
\label{eq:xyFunctionXiEta}
\begin{cases}
x  =  e^\xi (\xi \cos \eta - \eta \sin \eta)\\
y  =  e^\xi (\eta \cos \eta + \xi \sin \eta )\, .
\end{cases}
\end{equation}

Let us introduce $x_c=\tau D \Lambda^{\bar{\alpha}} $ and $\xi_c=-\tau A$, values for which the equality holds in~\eqref{eq:tosolve}. Observe that $x_c<0$ and $\xi_c>0$. Rewriting Eq.~\eqref{eq:xyFunctionXiEta} for $z_c=x_c$ (namely $y_c=0$) and $w_c=\xi_c+i\eta_c$ we get:
\begin{equation*}
\begin{cases}
x_c  =  e^{\xi_c} (\xi_c \cos \eta_c - \eta_c \sin \eta_c)\\
0  =  e^{\xi_c} (\eta_c \cos \eta_c + \xi_c \sin \eta_c )\, .
\end{cases}
\end{equation*}
From the second equation one can obtain (implicitly) $\eta_c$ as a function of $\xi_c$:
\begin{equation*}
{\eta_c} =-{\xi_c}\tan \eta_c\, .
\end{equation*}
Because $\xi_c<0$ one can find a unique solution $\eta_c(\xi_c)\in(\pi/2,\pi)$ (and the opposite one). Inserting this result into the first equation we get
\begin{equation*}
x_c(\xi_c)=- e^{-\eta_c(\xi_c)\mathrm{co}\!\tan \eta_c(\xi_c)} \frac{\eta_c(\xi_c)}{\sin\eta_c(\xi_c)} \, .
\end{equation*}

So in conclusion given $-\tau A$ one can obtain the critical $x_c(-\tau A)$ for which we have equality in Eq.~\eqref{eq:tosolve}, and thus conclude, by invoking the scaling properties of $W_0$, that for all $\Lambda^\alpha< x_c(-\tau A)/(\tau D)$ the strict inequality sign holds in Eq.~\eqref{eq:tosolve}.

\section{Turing-like instability are impeded when the delay only appears in the reaction term.}
\label{sec:delreact}
Let us consider system
\begin{equation}
\label{eq:main0}
\dot{x}_i(t)=f(x_i(t-\tau))+D \sum_j L_{ij} x_j(t)\, ,
\end{equation}
once the delay dependence is given only through the reaction term, $\tau_r=\tau>0$ and $\tau_d=0$. The condition for the stability of the homogeneous equilibrium without diffusion is given by $-\frac{\pi}{2}<\tau A<0$. Following the same procedure outlined in the main body of the paper -- linearising, then expanding the perturbation in the basis of the eigenvectors of the Laplacian -- one obtains the following characteristic equation:
\begin{equation}
\lambda_\alpha - A e^{-\lambda_\alpha\tau}-D\Lambda^\alpha =0\, ,
\end{equation}
that can be solved using the W--Lambert function to give\begin{equation}
\lambda_{\alpha,k} = D\Lambda^\alpha + \frac{1}{\tau}W_k(\tau A e^{-\tau D \Lambda^\alpha}),\quad k\in\mathbb{Z}\, .
\end{equation}

Turing-like instability can emerge if the homogeneous equilibrium becomes unstable in presence of the diffusion. We then look for $\alpha$ and $k$ such that $\Re \lambda_{\alpha,k}>0$. Using Lemma 3 of~\cite{ShinozakiMori2006} this is equivalent to
\begin{equation*}
\Re W_0 (\tau A e^{-\tau D \Lambda^\alpha})>-\tau D\Lambda^\alpha\, ,
\end{equation*}
Let us observe that $-\tau D\Lambda^\alpha\geq 0$. Because $\tau A<0$, one cannot solve the previous equation with $\Im W_0(\tau A e^{-\tau D \Lambda^\alpha} )=0$ (in this case one should have the argument of $W_0$ to be positive). Hence we look for $\tau Ae^{-\tau D \Lambda^\alpha}<-\pi/2$. 

Let us introduce $s=-D\Lambda^{\alpha}>0$, $u=\tau A\in(-\pi/2,0)$ and the function
\begin{equation}
\label{eq:cond20}
g(s)=-s+\Re W_0 (ue^s)\, ,
\end{equation}
our goal is to prove that $g(s)<0$ for all $s>0$ which is in turn equivalent to stating that Turing-like instability cannot develop.

Let us rewrite Eq.~\eqref{eq:xyFunctionXiEta} for $x=ue^s$ and $y=0$:
\begin{equation*}
\begin{cases}
ue^s  =  e^\xi (\xi \cos \eta - \eta \sin \eta)\\
0  =  e^\xi (\eta \cos \eta + \xi \sin \eta )\, .
\end{cases}
\end{equation*}
Isolating $\xi$ in the second equation and inserting it in the first one, we get:
\begin{equation*}
e^s=-\frac{\eta}{u\sin\eta}e^\xi\, .
\end{equation*}
One can thus rewrite $g(s)$ as follows:
\begin{eqnarray*}
g(s)&=&-s+\Re W_0 (ue^s)=-s+\xi=-\xi-\log\left(-\frac{\eta}{u\sin\eta}\right)+\xi \\
&=&-\log\left(-\frac{\eta}{u\sin\eta}\right)\, .
\end{eqnarray*}
Because $u>-\pi/2$ and $\pi/2< \eta<\pi$ we obtain $-\eta/(u\sin\eta)>1$ and thus $\log\left(-\frac{\eta}{u\sin\eta}\right)>0$.

\end{document}